\documentclass{article}

\usepackage{authblk}
\usepackage{graphicx}
\usepackage{amsfonts,amsmath}
\usepackage{siunitx}
\usepackage{csquotes}
\usepackage[sort]{cite}
\usepackage{natbib}
\usepackage{float}
\usepackage{hyperref}

\hypersetup{
    colorlinks,
    linkcolor={black},
    citecolor={black},
    urlcolor={black}
    }
\DeclareSIUnit\pixel{px}

\graphicspath{
  {/hri/storage/user/pwollsta/projects/interact/code/paid_data/results/results4paper/}
  {/hri/storage/user/pwollsta/projects/interact/figures/}
  {./figures/}}

\title{Quantifying the predictability of visual scanpaths using Active Information Storage}

\author[1,*]{Patricia Wollstadt}
\author[1]{Martina Hasenj{\"a}ger}
\author[1]{Christiane B. Wiebel-Herboth}

\affil[1]{Honda Research Insitute Europe GmbH, Carl-Legien-Str. 30, 63067 Offenbach/Main, Germany}
\affil[*]{Corresponding author: patricia.wollstadt@honda-ri.de}
\date{}

\begin{document}

\maketitle

\begin{abstract}
    Entropy-based measures are an important tool for studying human gaze
    behavior under various conditions. In particular, gaze transition entropy
    (GTE) is a popular method to quantify the predictability of fixation
    transitions. However, GTE does not account for temporal dependencies beyond
    two consecutive fixations and may thus underestimate a scanpath's actual
    predictability. Instead, we propose to quantify scanpath predictability by
    estimating the active information storage (AIS), which can account for
    dependencies spanning multiple fixations. AIS is calculated as the mutual
    information between a processes' multivariate past state and its next
    value. It is thus able to measure how much information a sequence of past
    fixations provides about the next fixation, hence covering a longer
    temporal horizon. Applying the proposed approach, we were able to
    distinguish between induced observer states based on estimated AIS,
    providing first evidence that AIS may be used in the inference of user
    states to improve human-machine interaction.
\end{abstract}


\section{Introduction}

The analysis of scanpaths has gained renewed interest in recent years, for
example, to study cognitive function \cite{Hayes2017,Raptis2017}, personality
traits \cite{Baranes2015,Allsop2017}, or as marker in gaze-based applications
\cite{Ebeid2018}. In particular, information-theoretic measures have become a
popular tool for studying cognitive function through the analysis of human gaze
behavior
\cite{Shiferaw2019,Hao2019,DiStasi2016,Shiferaw2018,Schieber2008,Krejtz2014,Krejtz2015,Raptis2017}.
A commonly used measure is the \textit{(gaze) transition entropy} (GTE)
\cite{Krejtz2015}, which uses a conditional Shannon Entropy
\cite{Shannon1948} to describe the regularity of transitions between fixations
\cite{Shiferaw2019}. GTE considers sequences of fixations, so-called scanpaths,
under the assumption that scanpaths can be modeled as Markov chains of order
one, and is calculated as the entropy of the transitions between two
consecutive fixations.
Low GTE---and thus a low remaining uncertainty about the location of
the next fixation given the previous one---thereby are interpreted as a high
predictability of the next fixation \cite{Shiferaw2019}. GTE has been applied
in various studies (see \cite{Shiferaw2019} for a review), which have shown
that changes in GTE are associated with higher task demand
\cite{DiStasi2016,Chanijani2016,Diaz2019}, increased anxiety
\cite{Allsop2014,Dijk2011,Gotardi2018}, or sleep deprivation
\cite{Shiferaw2018}.

Despite the popularity of information-theoretic measures, alternative
approaches have been employed in the analysis of scanpaths (e.g.,
\cite{Simola2008,Hayes2011,Kuebler2017,Coutrot2018}). A number of these studies
have found evidence for the importance of incorporating long-range temporal
information when analyzing and modeling eye movement data
\cite{Hayes2011,Hayes2017,Wiebel-Herboth2020}.

Hayes et al. \cite{Hayes2011}, for example, introduced the Successor
Representation (SR model) to the analysis of scan pattern, which uses an
algorithm from reinforcement learning to represent sequential gaze data in a
temporally extended fashion. More precisely, the SR model stems from temporal
difference learning~\cite{Sutton1988,Dayan1993} and incorporates among others a
temporal parameter that defines the time span for which an observation
(fixation) influences the model outcome. The authors found that up to
\SI{40}{\percent} of the variance in viewer intelligence, working memory
capacities, and speed of processing could be explained based on differences in
scan patterns that were individually modelled with the SR model
~\cite{Hayes2011, Hayes2017}, as well as  to some extent variances in ADHD
scores (up to \SI{50}{\percent}), autism quotients (up to \SI{30}{\percent})
and dyslexia scores (up to \SI{25}{\percent}) \cite{Hayes2018}. Wiebel-Herboth
et al. \cite{Wiebel-Herboth2020} found that an SR model had a significantly
higher predictive power when classifying single participants based on their
scan pattern in a visual comparison task compared to a simple transition matrix
model that considered only the immediate last fixation. Moreover, Hoppe et al.
\cite{Hoppe2019} were the first to provide quantitative evidence that humans
are capable of planning eye movements beyond the next fixation. Taken together,
these results suggest that longer temporal dependencies in scan pattern might
be informative about their underlying cognitive processes and thus should
be included in the modeling process.

Yet, entropy-based measures commonly applied, e.g. the GTE
\cite{Krejtz2014,Krejtz2015,Shiferaw2019}, typically only take into account
information contained within the immediate past fixation when quantifying the
regularity of eye movements. When GTE was firstly introduced, Krejtz and
colleagues \cite{Krejtz2014,Krejtz2015} adopted the procedure by
\cite{Besag2013} for testing the Markov chain of order one assumption underlying
the GTE computation. In their experiment, they found that in most cases the
assumption was valid, yet not in all. To our knowledge, such validation
procedure has however not become a standard procedure in the entropy-based gaze
analysis literature (for a review see \cite{Shiferaw2019}).
In cases where the order-one Markov chain assumption is violated,
longer temporal dependencies are not accounted for in current
information-theoretic approaches to scan path analysis. As a result, if such
temporal dependencies existed in a scanpath, the GTE would presumably
underestimate its overall predictability.

Alternative modeling approaches, such as the SR model, also come
with drawbacks. Most importantly, the model parameters have to be defined
ad-hoc and cannot be learned in a data driven fashion. This entails the risk of
a circular argumentation if no external optimization criterion can be defined.
Furthermore, the model parameters are not interpretable in a straightforward
way, which limits the explanatory power of the approach. Thus, there is still a
need for methods of scanpath modeling that can integrate both spatial and
temporal information in the data \cite{Krol2019}.

To this end, we here propose a novel approach to the
in\-for\-mation\--theo\-retic analysis of scanpath data, which is able to
measure predictability in a scanpath while accounting for temporal dependencies
of arbitrary order: we propose to estimate active information storage (AIS)
\cite{Lizier2012} from scanpaths, which measures the predictability of a
sequence as the mutual information between the sequences' past and its next
state. In particular, the relevant past is modeled as the collection of all
past variables that provide information about the next value, and can be
identified using novel estimation procedures that optimize the past state in a
data-driven fashion \cite{Faes2011,Wollstadt2019,Novelli2019}.

AIS has been successfully  applied in a variety of disciplines to measure
predictability of time series
\cite{Gomez2014,Brodski2017,Wollstadt2017,Faes2013,Wang2012,Lizier2012c}. In
the context of scanpath analysis, we believe AIS can provide several benefits
at once: 1) it implicitly tests the order-one Markov chain assumption, as it
provides the optimized past state for a given data sample. As such it can
provide direct evidence for whether fixations beyond the last fixation have a
predictive value. 2) The length of the optimized past state is directly
interpretable. That is, the optimization finds the temporal horizon over which
past fixation(s) are informative about and thus has the potential to support
the generation of explanatory hypothesis. 3) AIS allows for an individually
optimized computation of predictability that may be of greater usefulness in
gaze-based applications, e.g., driver assistance
\cite{Ebeid2018,Shiferaw2018,Shiferaw2019alcohol}. In sum, we argue that AIS
may be applied to quantify the predictability of a scanpath, in particular,
while information provided by fixations beyond the immediate fixation can be
detected and accounted for.

In the following, we will introduce AIS together with the necessary
information-theoretic background and describe its estimation from scanpath
data; as a proof of concept, we estimate AIS from scanpath data recorded in a
visual comparison task and show how variations in AIS reflect differences in
induced observer states.

\section{Materials and Methods}

\subsection{Information-theoretic preliminaries}

Formally, we consider a scanpath as realizations $(x_1, \ldots,x_t, \ldots,
$ $x_N)$, $x_t \in \mathcal{A}_{X_t}$ of a stationary random process
$\mathcal{X}=(X_1, X_2, \ldots, X_t,$ $\ldots, X_N)$, where a random process is
a collection of random variables, $X$, ordered by an integer $t \in \left\{1,
\ldots, N\right\} \subseteq \mathbb{N}$. As a shorthand, we write $p(x_t)$ for
the probability, $p(X_t=x_t)$, of variable $X_t$ taking on the value $x_t \in
\mathcal{A}_{X_t}$, and $p(x_t|y)$ for the conditional probability of $X_t$
taking on the value $x_t$ if the outcome $y$ of a second variable, $Y$, is
known.

The Shannon entropy \cite{Shannon1948} is then defined as

\begin{equation}
  H(X) = - \sum_{x \in \mathcal{A}_X} p(x) \log p(x)
  \label{eq:entropy}
\end{equation}

\noindent and quantifies the expected uncertainty associated with the random
variable $X$, or the amount of information to be gained when observing outcomes
of $X$. The conditional entropy is then the average information or uncertainty
remaining in $X$ if the outcome of $Y$ is known:

\begin{equation}
  H(X|Y) = - \sum_{x \in \mathcal{A}_X, y \in \mathcal{A}_Y} p(x,y) \log p(x|y).
  \label{eq:cond_entropy}
\end{equation}

Based on these definitions, we define the mutual information (MI) as the
average amount of information one variable, $X$, provides about a second
variable, $Y$,

\begin{equation}
  \begin{aligned}
    I(X;Y) =& H(X) - H(X|Y) = H(Y) - H(Y|X) \\
           =&  \sum_{x,y} p(x,y) \log \frac{p(x|y)}{p(x)}.
    \label{eq:mi}
  \end{aligned}
\end{equation}

\begin{figure}[h]
  \centering
  \makebox[\textwidth][c]{\includegraphics[width=1.3\textwidth]{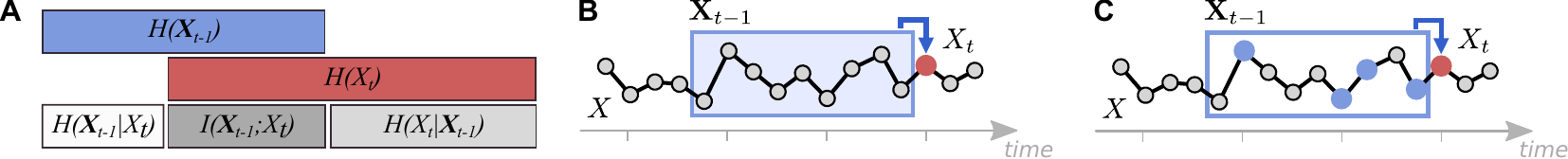}}%
  \caption{
    (A) Relationship between entropy and conditional entropy, $H$, and mutual
    information $I$ of two non-independent random variables $X_t$ and
    $\mathbf{X}_{t-1}^-$ (adapted from \protect\cite{MacKay2005}). Here, the
    conditional entropy corresponds to the gaze transition entropy (GTE).
    (B) Active information storage (AIS) quantifies the predictability of the value
    of time series $X$ at time $t$, $X_t$ (red marker), from its immediate
    past state, $\mathbf{X}_{t-1}^-$ (blue box).
    (C) Non-uniform embedding representing the past state of time series, $X$,
    as a selection of past variables (blue markers) up to a maximum lag
    $k_{max}<t$, which carry significant information about the next value,
    $X_t$ (red marker).}
  \label{fig:ais_intro}
\end{figure}

\noindent The MI quantifies the information $X$ provides about $Y$ and
\textit{vice versa}; it is zero for independent variables ($p(x,y)=p(x)p(y)$)
or if either $H(X)$ or $H(Y)$ are zero, i.e., there is no information to share.
The MI is bound from above by the entropy of both variables involved, $0 \leq
I(X;Y) \leq H(X), H(Y)$ (Fig. \ref{fig:ais_intro}A).

\subsection{Active Information Storage (AIS)}

AIS \cite{Lizier2012} quantifies how much information a processes' past state
$\mathbf{X}_{t-1}^-$ contains about its next value $X_{t}$ and thus measures
the average predictability of $X_t$ from its immediate past
\cite{Lizier2012,Wibral2014} (Fig. \ref{fig:ais_intro}B). AIS is calculated as
the MI between $\mathbf{X}_{t-1}^-$ and $X_t$,

\begin{equation}
  \begin{aligned}
    AIS(X_t) &= I(\mathbf{X}_{t-1}^-; X_t) = H(X_t) - H(X|\mathbf{X}_{t-1}^-) \\
          &= \sum_{x_t, \mathbf{x}_{t-1}^-}
              p(x_t, \mathbf{x}_{t-1}^-) \log \frac{p(x_t|\mathbf{x}_{t-1}^-)}{p(x_t)}, \\
    \label{eq:ais}
  \end{aligned}
\end{equation}

\noindent where the past state $\mathbf{X}_{t-1}^-$ is defined as a collection
of random variables up to a maximum lag $k_{max}$ (see also next section),

\begin{equation}
    \mathbf{X}_{t-1}^- = \left\{X_{t-1}, \ldots, X_{t-t_l}, \ldots, X_{t-k_{max}} \right\}.
\end{equation}

AIS is low for processes with highly random transitions and high for processes
that visit many different states in a regular fashion
\cite{Wibral2014,Lizier2012}. Formally, $0 \leq AIS(X_t) \leq H(X_t),
H(\mathbf{X}_{t-1}^-)$, i.e., AIS is zero for processes with no memory such
that they are completely random, and the AIS is upper bounded by the entropy of
the past state and entropy of the next value of a process.

\subsubsection{Relationship between GTE and AIS}

GTE measures the remaining uncertainty in a fixation, given knowledge of the
previous fixation as a conditional entropy, $H(X_t|X_{t-1})$. Hence, for past
states of length one, AIS and GTE are complementary, i.e., $H(X_t) =
I(X_t;X_{t-1}) + H(X_t|X_{t-1})$ (eq. \ref{eq:ais} and
Fig.\ref{fig:ais_intro}A). However, for processes that do not fulfill the
Markov condition, $p(X_t|X_{t-1}, \ldots, X_{t-l}) = p(X_t|X_{t-1})$, i.e.,
processes that are not sufficiently described by a Markov chain of order one,
the GTE may underestimate the actual predictability of the next state from the
whole \textit{relevant} past of $X$ (see also next section). Furthermore, both
measures differ in their interpretation---while the GTE measures the remaining
uncertainty in the next fixation, the AIS as a MI, measures how much
information the past provides about the next fixation. The latter thus provides
a more direct measure of predictability \cite{Wibral2014,Crutchfield2003}.

\subsection{Estimating AIS from scanpath data}

\subsubsection{Optimization of past states}

To estimate AIS in practice, we first have to define the past state
$\mathbf{X}_{t-1}^-$ such that it contains all relevant information stored in
the past of $X$ about $X_t$ \cite{Lizier2012}. Formally, we want to define
$\mathbf{X}_{t-1}^-$ such that $p(X_t|X_{t-1}, \ldots, X_{t-l}) =
p(X_t|\mathbf{X}_{t-1}^-)$. In other words, the next value, $X_t$, is
conditionally independent of all past variables, $X_{t-l}, l > k_{max}$, given
$\mathbf{X}_{t-1}^-$. Non-optimal choices for $\mathbf{X}_{t-1}^-$ may lead to
an underestimation of AIS if not all relevant information is covered by
$\mathbf{X}_{t-1}^-$, or they may lead to artificially inflated AIS values if
too many variables are included, leading to an under-sampling of the past
state.

Here, we find $\mathbf{X}_{t-1}^-$ through a non-uniform embedding procedure
\cite{Faes2011, Lizier2012b} that selects a subset of variables from all past
variables up to a maximum lag, $k_{max}$ (Fig. \ref{fig:ais_intro}C),

\begin{equation}
  \mathbf{X}_{t-1}^- = \left\{X_{t-k}\right\},\,  k \in \left[1, k_{max}\right].
  \label{eq:nonuniform_embedding}
\end{equation}

We optimize $\mathbf{X}_{t-1}^-$ using a greedy forward-selection approach
implemented in \cite{Wollstadt2019}, which iteratively includes variables if
they provide significant, additional information about $X_t$, conditional on
all already selected variables. The implementation uses a hierarchical
permutation testing scheme to handle estimator bias, while controlling the
family-wise error rate during repeated testing \cite{Novelli2019}. Using
statistical testing for the inclusion of variables further provides an
automatic stopping criterion for construction of the past state.



\subsubsection{Estimating AIS from discrete scanpath data}

After optimizing $\mathbf{X}_{t-1}^-$, we estimate AIS from scanpath data using
plug-in estimators \cite{Hlavackova-Schindler2007} that are known to exhibit a
bias due to finite sampling (e.g., \cite{Paninski2003,Miller1955}). Our
approach to handling estimator bias is two-fold: first, we apply a
bias-correction proposed in \cite{Panzeri1996,Panzeri2007} and implemented in
\cite{Ince2009} to final AIS- and entropy-estimates; second, we use
non-parametric permutation testing \cite{Vicente2011,Novelli2019} to test
MI-estimates for statistical significance.

We here use permutation testing during the optimization as well as to
test final AIS estimates. Permutation testing considers the MI-estimate as a
test statistic in a test against the null-hypothesis of no relationship between
the two variables, where the null-distribution is found through repeated
estimation from permuted data \cite{Vicente2011}.

\subsection{Experiment}

As a proof of concept, we estimate AIS from eye tracking data recorded during a
visual comparison task, where we varied observer states by adding a time
constraint in one condition (see also \cite{Wiebel-Herboth2020} for details on
the experimental setup).

\subsubsection{Participants}

We recorded data from \num{13} participants (one female) with a mean age of
\num{38}, ranging from \numrange{21}{53}. Data of three participants had to be
excluded from the analysis due to failures in the recording process,
resulting in a sample of \num{10} all-male participants. All participants had
normal or corrected to normal sight and gave their informed written consent
before participating in the experiment.

\subsubsection{Task and experimental procedure}

Participants were asked to identify as fast as possible the difference between
a reference and a target image, where both images were identical except for one
detail that was changed in the target image. Both images were presented next to
each other on a mean gray background (Fig.~\ref{fig:procedure}A). Participants
were asked to indicate the location of the difference by clicking, using a
regular computer mouse. The experiment took place in a quiet office environment
under normal lighting conditions. Before the start of the experiment,
participants were informed about the course of the experiment and received
instructions.

Trials were recorded under two experimental conditions that were
designed such as to induce two different user states, one relaxed state and one
in which participants experienced stress through time pressure. Time pressure
was achieved by varying the time available for the participants to complete the
search task in each trial: in the first condition, participants had as much
time as they needed (time unconstrained condition, TUC); in the second
condition, time to finish the task was constraint to \SI{9}{\second} (time
constrained condition, TC). The time limit was chosen such that it would lead
to a significant performance drop and was determined in pre-tests. In addition,
a sequence of nine accelerating tones, presented via headphones indicated the
time running up in the TC condition. If participants did not find the
difference between the images within the given time range, the next trial was
initiated independently of the participant's response. Performance dropped on
average to \SI{65}{\percent} correct trials in the TC condition with an average
search time of $m= 4.84 s$, $sem =0.15s$, (TUC: \SI{100}{\percent}, average
search time: $m= 17.01s$, $sem = 1.5s$). This result indicates that the intended
manipulation was indeed successful. All participants reported after the
experiment that they felt under time pressure in the TC condition.
For each condition \num{22} trials were recorded. After half of the trials,
participants were asked to take a break.

\subsubsection{Apparatus and stimuli}

Stimuli were presented on a Dell monitor. Participants saw \num{44} photographs
of varying indoor and outdoor scenes. Images were taken from a publicly
available database
(\href{http://search.bwh.harvard.edu/new/Shuffle_Images.html}{Shuffle database,
Large Change Images}) \cite{Sareen2016}. The experimental routine was
programmed in Python using Psychopy~\cite{Peirce2007,Peirce2009}. The
participants' gaze behavior was recorded using a pupil labs eye tracker, using
\SI{120}{Hz} binocular gaze tracking and \SI{60}{fps} world camera recordings
\cite{Kassner2014}.

\begin{figure}
  \centering
  \includegraphics[width=1.0\linewidth]{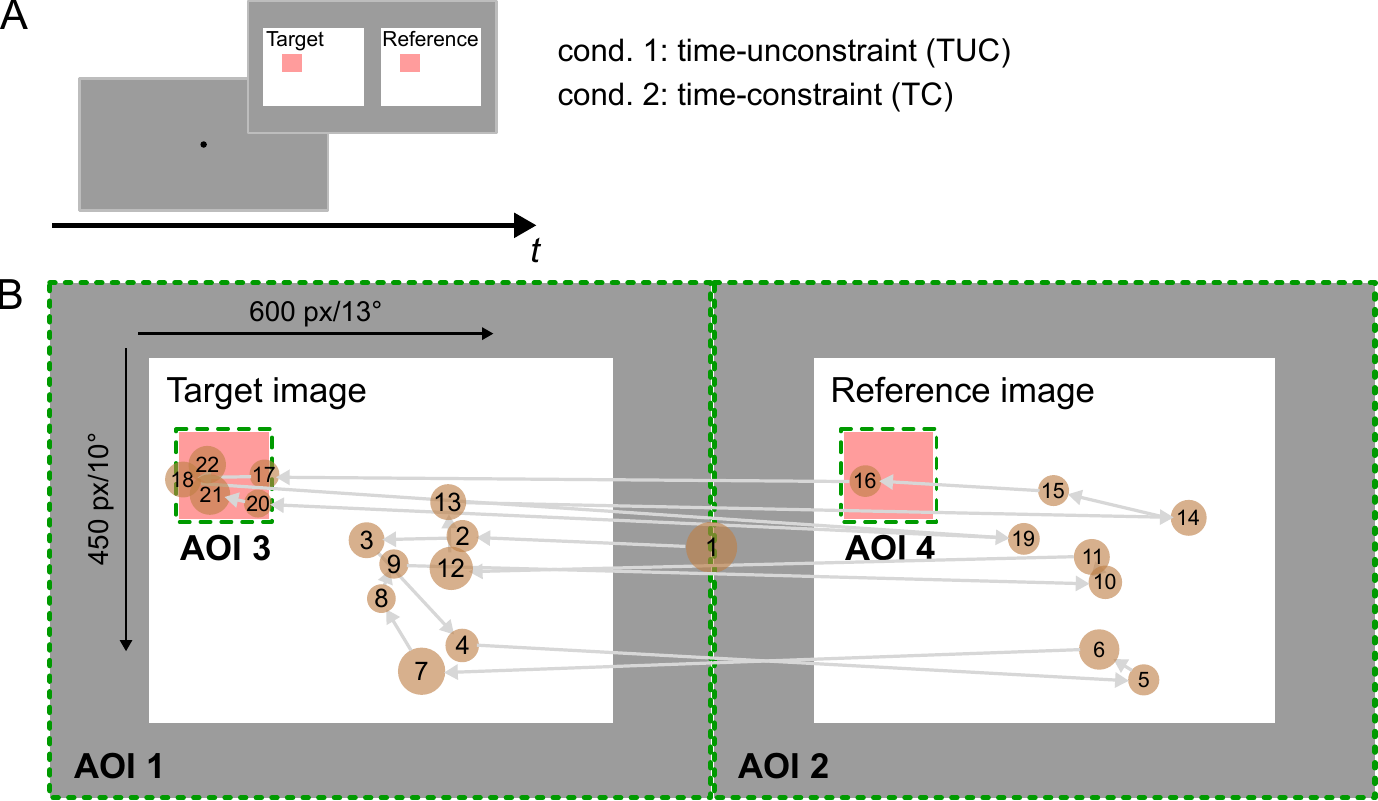}
  \caption{(A) Experimental setup for individual trial consisting of a screen
    showing the fixation dot and a screen displaying the image pair on a mean
    grey background.
    (B) Definition of areas of interest (AOI, green boundaries) on schematic
    images with target area (red box). The white line denotes an exemplary
    scanpath, where orange markers indicate ordered fixations and marker size
    corresponds to fixation time.}
  \label{fig:procedure}
\end{figure}

The eye tracker was calibrated at the beginning of the experiment. All
calibrations were done using the 9-point calibration routine implemented by
pupil labs. Gaze points were mapped to the screen via the screen marker
solution implemented by pupil labs. For that purpose the monitor was defined as
a surface based on \num{10} markers attached to the edge of the screen. To
validate the calibration, participants were asked to fixate on a fixation dot
presented at the center of the screen at the beginning of each trial
(Fig.~\ref{fig:procedure}A). The pupil labs eye tracker offers an accuracy of
up to \SI{0.6}{\degree} and a precision of \SI{0.2}{\degree}. If
online-computed deviations between the recorded gaze position and the fixation
dot exceeded \SI{50}{\pixel} (corresponding to a viewing angle of \ang{1.15}),
the eye-tracker was recalibrated.

Images presented during a trial categorized into easy, medium and
difficult with respect to the search task prior to the experiment. Ratings were
done by three experimenters independently resulting in \SI{75}{\percent} of all
ratings to be \SI{100}{\percent} consistent whereas for the remaining
\SI{25}{\percent} (eleven images), ratings deviated by one (e.g.: easy, easy,
medium). To resolve these inconsistent cases, the median of the ratings was
chosen as a label (e.g.: easy). The image dataset was split in half assuring an
equal distribution of difficulty among the two. Half of the images were used
for the TC condition ($n=22$) while the other half was used for the TUC
condition ($n=22$). Within each condition, images were shown in a randomized
order, with no image shown twice to the same observer.

\subsubsection{Preprocessing}

Data analysis was done in Python and R \cite{R2017}. Fixations for scanpath
representations were computed using the basic Identification by
Dispersion-Threshold (IDT) algorithm~\cite{Salvucci2000} using a maximum
dispersion of \SI{50}{\pixel} and a minimum duration of
\SI{100}{\milli\second}. Fixations above \SI{1500}{\milli\second} and data
points with a confidence value below \num{0.9} were excluded from the data
analysis. We analyzed data from all trials and did not differentiate
between correct and incorrect trials.

Scanpaths were defined as sequential fixations of predefined areas of interest
(AOI) and thus represent time-series data incorporating temporal as well as
spatial information. We defined AOIs as four areas of interest
(Fig.~\ref{fig:procedure}B): (1) the left half of the monitor, (2) the right
half of the monitor, (3) the target area in the left image and (4) the
respective target area in the right image. Target areas were defined based on
the bounding boxes specifying the location of difference plus an additional
frame of \SI{50}{\pixel}. Our approach aimed at extracting differences in the
search process related to a presumably first \enquote{general search phase} and
a \enquote{zooming in and validating phase} at the end of each trial.

\section{Results}

\subsection{Optimization of past states}

\begin{figure}[h]
  \centering
  \includegraphics[width=0.9\linewidth]{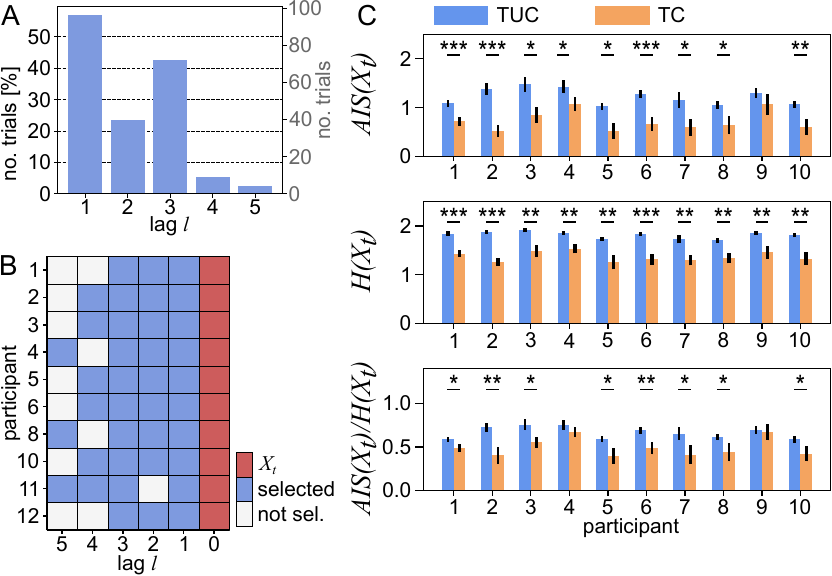}
  \caption{
    (A) Number of past variables  with a given lag $l$ selected through
    non-uniform embedding over participants and trials.
    (B) Union past state for each participant used for statistical testing.
    Selected past variables (blue) with lag $l$ relative to the next fixation,
    $X_t$ (red).
    (C) Mean active information storage (AIS, top), entropy (middle), and
    normalized AIS (bottom), for conditions, \textit{time constraint} (TC) and
    \textit{time unconstrained} (TUC), and individual participants
    (*$p<0.05$, **$p<0.01$, ***$p<0.001$, error bars indicate the standard
    error of the mean).}
  \label{fig:results}
\end{figure}

We estimated AIS from scanpaths for each trial individually using the IDTxl
python toolbox \cite{Wollstadt2019}. We first optimized past states,
$\mathbf{X}_{t-1}^-$, while setting $k_{max}$ to \num{5} previous fixations.
This resulted in a wide variety of selected past variables over trials and
participants, where in \SI{74}{\percent} of trials, variables with lags greater
one were selected (Fig. \ref{fig:results}A). Hence, in the majority of trials,
fixations prior to the last fixation provided significant information about the
next fixation and were relevant for quantifying the predictability of the
scanpath. Furthermore, the variability in lags provides evidence for an intra-
and inter-individual variance in viewing behavior that should be accounted for
by estimation procedures.

\subsection{Difference in experimental conditions}

\subsubsection{Overall effect of condition on predictability}

To test for significant differences in predictability between the two
experimental conditions, we fitted a linear mixed effects model with fixed
effect \textit{experimental condition} and random effect \textit{participant},
allowing for a varying random slope for the effect of experimental condition on
AIS values per participant \cite{Fahrmeir2007}. For fitting the model, we used
the lme4 package \cite{Bates2015}, written in R \cite{R2017}.

We found a main effect of experimental condition ($\chi^2(1) = 30.054$,
$p<0.001$), while we found no significant effect of the random slope. This
indicates an overall effect of experimental condition on predictability when
controlling for inter-subject variability. To assess how the experimental
condition affected predictability for individual participants, we performed for
each participant an independent samples permutation test between AIS in both
conditions ($N_{perm}=5000$, Fig. \ref{fig:results}C). We found significantly
decreased AIS in nine out of ten participants in the TC condition ($p<0.05$, $
AIS_{TC}(X_t) < AIS_{TUC}(X_t)$).

\subsubsection{Relationship between AIS and scanpath entropy}

In a second step, we investigated whether the decrease in AIS reflected a true
decline in the predictability of the scanpath or was rather due to a lower
scanpath entropy in the TC condition. Since the absolute AIS value is bounded
by the entropy of the two variables involved, a reduction in \textit{absolute}
AIS may not only be caused by change in the predictability of a process, but
also by a reduction in the processes' entropy, i.e., a reduction in the
information to be predicted.

We performed two-tailed, independent samples per\-mu\-ta\-tion test for
differences in  $H(X_t)$ between conditions for each participant
($N_{perm}=5000$, Fig. \ref{fig:results}), where we found a significant decline
in  $H(X_t)$ for the TC condition for all participants ($p<0.05$, $H_{TC}(X_t)
< H_{TUC}(X_t) $). To investigate if the decrease in $H(X_t)$ may fully explain
the decrease in $AIS(X_t)$, we further tested for differences in AIS normalized
by $H(X_t)$, $AIS(X_t) / H(X_t)$. Here we found a significant decline in eight
out of ten participants ($p<0.05$, $AIS_{TC}(X_t) / H_{TC}(X_t) <
AIS_{TUC}(X_t) / H_{TUC}(X_t)$). This result indicated that the observed
decrease in absolute AIS may be at least partially explained by a reduction in
fixation entropy, but also by an actual reduction in the regularity or
predictability of the scanpath.

Note that for all statistical comparisons, to avoid spurious effects, we aimed
at holding estimation bias constant between the groups compared. Estimation
bias depends on the number of samples and the size of the variables used
\cite{Panzeri2007,Paninski2003}. Hence, we fixed the number of samples by
discarding samples at the beginning of a trial and created a uniform past state
by taking the union of selected past states over all trials and conditions
(Fig. \ref{fig:results}B). Taking the union ensures that the uniform
past state contains all relevant variables at the expense of including
potentially irrelevant variables in the estimation from some of the trials.


\section{Discussion}

We presented AIS \cite{Lizier2012} as a novel approach to quantifying the
predictability of scanpaths while accounting for long-range temporal
dependencies between fixations. We demonstrated how to estimate AIS from
scanpath data recorded during a visual comparison task and found that changes
in viewer states were reflected by changes in estimated AIS, indicating a lower
predictability of gaze behavior in more demanding task conditions.

Current information-theoretic measures of predictability in scanpaths do not
incorporate long-range temporal information, which may be important to
accurately describe human viewing behavior
\cite{Hayes2011,Hayes2017,Hayes2018,Hoppe2019,Wiebel-Herboth2020}. Alternative
measures, such as the SR model \cite{Hayes2011} or hidden Markov models
\cite{Simola2008,Coutrot2018} lack in interpretability and their application is
not always straightforward \cite{Hayes2011,Shiferaw2019}. For example, the
learning parameter representing the temporal horizon in the SR model has no
clear interpretation, such that it is typically set through optimization of an
additional criterion. Such an external criterion may not be readily available
and may lead to circular analysis designs \cite{Kriegeskorte2009}.

In contrast, AIS paired with novel estimation techniques, namely non-uniform
embedding using a recently proposed estimation algorithm
\cite{Faes2011,Wollstadt2019,Novelli2019}, allows to optimize the temporal
horizon accounted for in a purely data-driven fashion. Furthermore, the
optimized past state allows for a clear interpretation in units of past
samples, offering additional explanatory value. Lastly, the past state is
optimized individually per participant, accounting for inter-individual
variation and including cases that are best modeled by a Markov chain of order
one as a special case. In the latter case, i.e., if the optimized past state
contains only the past fixation with lag one, AIS and GTE are complementary
such that a change in GTE corresponds to an equivalent change in AIS and vice
versa. When applying AIS estimation to scanpath data, we found significant
temporal relationships in scanpaths beyond first-order transitions and high
inter-individual variability. Both findings underline the importance of
accounting for long-range temporal dependencies as well as inter-individual
differences when modeling scanpath data, in particular when quantifying the
regularity or predictability of gaze behavior.




As a proof of concept we applied the AIS estimation to eye tracking data
recorded from a visual comparison task, in which two different observer states
were induced. In the TC condition, participants experienced a higher task
demand compared to the TUC condition. Here, we found a significant decline in
predictability measured by AIS for higher task demand. This result is in line
with the majority of studies utilizing GTE, which find an increase in GTE and
thus lower predictability under increased task difficulty (see
\cite{Shiferaw2019} for a review). We conclude that AIS is able to detect
changes in predictability due to changes in task demand, while avoiding
methodological \textit{ad hoc} choices and being more versatile with respect to
temporal correlations present in gaze behavior. Furthermore, we argue that AIS
provides a more immediate measure of \enquote{predictability} as its
calculation incorporates the \textit{maximum} amount of information that can be
gained from the past of a process about its next state
\cite{Crutchfield2003,Wibral2014}. Lastly, note that we here normalized our
estimate of predictability by the entropy of the next fixation, also termed
\textit{stationary gaze entropy} \cite{Shiferaw2019}. We emphasize that such a
normalization is necessary to exclude that changes in GTE or AIS are purely due
to a change in fixation entropy between experimental conditions.

Being able to quantify changes in user states using approaches such
as the one presented here, is central, for example, in many human machine
cooperation scenarios \cite{Krueger2017}. Being able to detect user states and
changes therein, allows to adapt machine behavior such as to improve the
interaction. Imagine for example a teaching assistance system, which in order
to provide optimal support for a student, must be able to assess whether a
change in task demand, e.g., increasing the level of difficulty, is appropriate
or overextending for the student. Only then the system can adjust to the right
level of information supply or offer additional support for solving the task
(see for example \cite{Celiktutan2018}). For such an assessment of the human
state, gaze behavior has been suggested as a rich data source, whose analysis
can provide unobtrusive insights into a user's cognitive or emotional state
(e.g.: \cite{Zagermann2018,Majaranta2014,Shiferaw2019}). For the analysis of
gaze behavior, in particular information-theoretic measures have been suggested
as promising markers of human states. We here extend existing work in this
field by including previously neglected temporal correlations in the analysis
of scanpath predictability, and thereby forego a potential underestimation and
thus misinterpretation of user states. We therefore suggest AIS as an novel
information-theoretic approach to the analysis of gaze behavior in user state
estimation.

\section{Conclusions}

We conclude that AIS is a promising measure for analyzing the predictability of
scanpath data. Future work should extend its application to eye tracking data,
for example, by exploiting the possibility to interpret AIS in a local
(sample-wise) fashion \cite{Lizier2012}. Information-theoretic quantities such
as entropy or mutual information allow for an interpretation for individual
realizations of the random variables involved, which allows to quantify the
local entropy or local predictability of a single fixation in time. Such a
localized description of fixation sequences allows for a more fine-grained
quantification of gaze behavior, up to the quantification of the predictability
of a single fixation. Applying AIS in its localized version thus opens the
possibility of using information-theoretic measures in real-time applications
such as online monitoring or assistance (e.g. \cite{Ebeid2018}).
Furthermore, the application should be extended to other tasks, in particular free viewing, to gain further insights on how predictability changes as a function of the task at hand and given more natural viewing conditions. This will be a next important step to evaluate its potential for real-world gaze-based applications. Also, note that our study is limited by its relatively small sample size and an all-male sample. Future research should therefore extend the application of AIS to larger and more diverse groups. Lastly,
AIS estimation may be applied to non-discretized fixation coordinates by
foregoing the definition of AOIs and applying estimators for continuous data
\cite{Kraskov2004,Khan2007}. Further studies may explore these possibilities to
further evaluate and extend the application of AIS to scanpath data.

\section*{Glossary}
\begin{tabular}{@{}ll}
AIS & Active Information Storage\\
GTE & Gaze Transition Entropy\\
IDTxl & Information Dynamics Toolkit xl\\
IDT & Identification by Dispersion-Threshold \\
MI & Mutual Information \\
SR Model & Successor Representation Model
\end{tabular}

\bibliographystyle{plain}
\bibliography{ais_scan_path}

\begin{thebibliography}{}

\bibitem[Allsop and Gray, 2014]{Allsop2014}
Allsop, J. and Gray, R. (2014).
\newblock Flying under pressure: Effects of anxiety on attention and gaze
  behavior in aviation.
\newblock {\em Journal of Applied Research in Memory and Cognition},
  3(2):63--71.

\bibitem[Allsop et~al., 2017]{Allsop2017}
Allsop, J., Gray, R., B{\"{u}}lthoff, H.~H., and Chuang, L. (2017).
\newblock {Effects of Anxiety and cognitive load on instrument scanning
  behavior in a flight simulation}.
\newblock In {\em Proceedings of the 2nd Workshop on Eye Tracking and
  Visualization (ETVIS) 2016}, pages 55--59, New York, NY. {IEEE}.

\bibitem[Baranes et~al., 2015]{Baranes2015}
Baranes, A., Oudeyer, P.-Y., and Gottlieb, J. (2015).
\newblock Eye movements reveal epistemic curiosity in human observers.
\newblock {\em Vision Research}, 117:81--90.

\bibitem[Bates et~al., 2015]{Bates2015}
Bates, D., M{\"a}chler, M., Bolker, B., and Walker, S. (2015).
\newblock Fitting linear mixed-effects models using {lme4}.
\newblock {\em Journal of Statistical Software}, 67(1):1--48.

\bibitem[Brodski-Guerniero et~al., 2017]{Brodski2017}
Brodski-Guerniero, A., Paasch, G.-F., Wollstadt, P., {\"O}zdemir, I., Lizier,
  J.~T., and Wibral, M. (2017).
\newblock Information-theoretic evidence for predictive coding in the
  face-processing system.
\newblock {\em Journal of Neuroscience}, 37(34):8273--8283.

\bibitem[Celiktutan and Demiris, 2018]{Celiktutan2018}
Celiktutan, O. and Demiris, Y. (2018).
\newblock Inferring human knowledgeability from eye gaze in mobile learning
  environments.
\newblock In {\em Proceedings of the European Conference on Computer Vision
  (ECCV)}.

\bibitem[Chanijani et~al., 2016]{Chanijani2016}
Chanijani, S. S.~M., Klein, P., Bukhari, S.~S., Kuhn, J., and Dengel, A.
  (2016).
\newblock Entropy based transition analysis of eye movement on physics
  representational competence.
\newblock In {\em Proceedings of the 2016 ACM International Joint Conference on
  Pervasive and Ubiquitous Computing (UbiComp): Adjunct}, pages 1027--1034, New
  York, NY. ACM.

\bibitem[Coutrot et~al., 2018]{Coutrot2018}
Coutrot, A., Hsiao, J.~H., and Chan, A.~B. (2018).
\newblock Scanpath modeling and classification with hidden {Markov} models.
\newblock {\em Behavior Research Methods}, 50(1):362--379.

\bibitem[Crutchfield and Feldman, 2003]{Crutchfield2003}
Crutchfield, J.~P. and Feldman, D.~P. (2003).
\newblock {Regularities unseen, randomness observed: Levels of entropy
  convergence}.
\newblock {\em Chaos}, 13(1):25--54.

\bibitem[Dayan, 1993]{Dayan1993}
Dayan, P. (1993).
\newblock Improving generalization for temporal difference learning: The
  successor representation.
\newblock {\em Neural Computation}, 5(4):613--624.

\bibitem[Di~Stasi et~al., 2016]{DiStasi2016}
Di~Stasi, L.~L., Diaz-Piedra, C., Rieiro, H., Carri{\'o}n, J. M.~S., Berrido,
  M.~M., Olivares, G., and Catena, A. (2016).
\newblock {Gaze entropy reflects surgical task load}.
\newblock {\em Surgical Endoscopy}, 30:5034--5043.

\bibitem[Diaz-Piedra et~al., 2019]{Diaz2019}
Diaz-Piedra, C., Rieiro, H., Cherino, A., Fuentes, L.~J., Catena, A., and
  Di~Stasi, L.~L. (2019).
\newblock The effects of flight complexity on gaze entropy: An experimental
  study with fighter pilots.
\newblock {\em Applied Ergonomics}, 77:92--99.

\bibitem[Ebeid and Gwizdka, 2018]{Ebeid2018}
Ebeid, I.~A. and Gwizdka, J. (2018).
\newblock Real-time gaze transition entropy.
\newblock In {\em Proceedings of the 2018 ACM Symposium on Eye Tracking
  Research \& Applications}, page Article No. 94, New York, NY, USA. ACM.

\bibitem[Faes et~al., 2011]{Faes2011}
Faes, L., Nollo, G., and Porta, A. (2011).
\newblock Information-based detection of nonlinear granger causality in
  multivariate processes via a nonuniform embedding technique.
\newblock {\em Physical Review E}, 5:051112.

\bibitem[Faes et~al., 2013]{Faes2013}
Faes, L., Porta, A., Rossato, G., Adami, A., Tonon, D., Corica, A., and Nollo,
  G. (2013).
\newblock Investigating the mechanisms of cardiovascular and cerebrovascular
  regulation in orthostatic syncope through an information decomposition
  strategy.
\newblock {\em Autonomic Neuroscience}, 178(1--2):76--82.

\bibitem[Fahrmeir et~al., 2007]{Fahrmeir2007}
Fahrmeir, L., Kneib, T., Lang, S., and Marx, B. (2007).
\newblock {\em Regression}.
\newblock Springer, Berlin, Heidelberg.

\bibitem[G{\'o}mez et~al., 2014]{Gomez2014}
G{\'o}mez, C., Lizier, J.~T., Schaum, M., Wollstadt, P., Gr{\"u}tzner, C.,
  Uhlhaas, P., Freitag, C.~M., Schlitt, S., B{\"o}lte, S., Hornero, R., et~al.
  (2014).
\newblock Reduced predictable information in brain signals in autism spectrum
  disorder.
\newblock {\em Frontiers in Neuroinformatics}, 8:9.

\bibitem[Gotardi et~al., 2018]{Gotardi2018}
Gotardi, G., Schor, P., Van Der~Kamp, J., Navarro, M., Orth, D., Savelsbergh,
  G., Polastri, P.~F., Oudejans, R., and Rodrigues, S.~T. (2018).
\newblock The influence of anxiety on visual entropy of experienced drivers.
\newblock In {\em Proceedings of the 3rd Workshop on Eye Tracking and
  Visualization (ETVIS)}, pages 1--4, New York, NY. ACM.

\bibitem[Hao et~al., 2019]{Hao2019}
Hao, Q., Sbert, M., and Ma, L. (2019).
\newblock Gaze information channel in cognitive comprehension of poster
  reading.
\newblock {\em Entropy}, 21(5):1--24.

\bibitem[Hayes and Henderson, 2017]{Hayes2017}
Hayes, T.~R. and Henderson, J.~M. (2017).
\newblock Scan patterns during real-world scene viewing predict individual
  differences in cognitive capacity.
\newblock {\em Journal of Vision}, 17(5):23,~1--17.

\bibitem[Hayes and Henderson, 2018]{Hayes2018}
Hayes, T.~R. and Henderson, J.~M. (2018).
\newblock Scan patterns during scene viewing predict individual differences in
  clinical traits in a normative sample.
\newblock {\em PLOS ONE}, 13(5):e0196654.

\bibitem[Hayes et~al., 2011]{Hayes2011}
Hayes, T.~R., Petrov, A.~A., and Sederberg, P.~B. (2011).
\newblock A novel method for analyzing sequential eye movements reveals
  strategic influence on raven's advanced progressive matrices.
\newblock {\em Journal of Vision}, 11(10):10.

\bibitem[Hlav{\'{a}}{\v{c}}kov{\'{a}}-Schindler et~al.,
  2007]{Hlavackova-Schindler2007}
Hlav{\'{a}}{\v{c}}kov{\'{a}}-Schindler, K., Palu{\v{s}}, M., Vejmelka, M., and
  Bhattacharya, J. (2007).
\newblock Causality detection based on information-theoretic approaches in time
  series analysis.
\newblock {\em Physics Reports}, 441(1):1--46.

\bibitem[Hoppe and Rothkopf, 2019]{Hoppe2019}
Hoppe, D. and Rothkopf, C.~A. (2019).
\newblock {Multi-step planning of eye movements in visual search}.
\newblock {\em Scientific Reports}, 9(1):1--12.

\bibitem[Ince et~al., 2009]{Ince2009}
Ince, R. A.~A., Petersen, R.~S., Swan, D.~C., and Panzeri, S. (2009).
\newblock Python for information theoretic analysis of neural data.
\newblock {\em Frontiers in Neuroinformatics}, 3:Article 4.

\bibitem[Julian and Mondal, 2013]{Besag2013}
Julian, B. and Mondal, D. (2013).
\newblock Exact goodness-of-fit tests for markov chains.
\newblock {\em Biometrics}, 69(2):488--496.

\bibitem[Kassner et~al., 2014]{Kassner2014}
Kassner, M., Patera, W., and Bulling, A. (2014).
\newblock Pupil: An open source platform for pervasive eye tracking and mobile
  gaze-based interaction.
\newblock In {\em In Proceedings of the 2014 ACM International Joint Conference
  on Pervasive and Ubiquitous Computing (UbiComp): Adjunct Publication}, pages
  1151--1160, New York, NY. ACM.

\bibitem[Khan et~al., 2007]{Khan2007}
Khan, S., Bandyopadhyay, S., Ganguly, A.~R., Saigal, S., Erickson~III, D.~J.,
  Protopopescu, V., and Ostrouchov, G. (2007).
\newblock Relative performance of mutual information estimation methods for
  quantifying the dependence among short and noisy data.
\newblock {\em Physical Review E}, 76(2):026209.

\bibitem[Kraskov et~al., 2004]{Kraskov2004}
Kraskov, A., St{\"{o}}gbauer, H., and Grassberger, P. (2004).
\newblock {Estimating mutual information}.
\newblock {\em Physical Review E}, 69(6):16.

\bibitem[Krejtz et~al., 2015]{Krejtz2015}
Krejtz, K., Duchowski, A., Szmidt, T., Krejtz, I., Perilli, F.~G., Pires, A.,
  Vilaro, A., and Villalobos, N. (2015).
\newblock Gaze transition entropy.
\newblock {\em ACM Transactions on Applied Perception}, 13(1):Article 4.

\bibitem[Krejtz et~al., 2014]{Krejtz2014}
Krejtz, K., Szmidt, T., Duchowski, A.~T., and Krejtz, I. (2014).
\newblock Entropy-based statistical analysis of eye movement transitions.
\newblock In {\em Proceedings of the Symposium on Eye Tracking Research and
  Applications (ETRA)}, pages 159--166, New York, NY. Association for Computing
  Machinery.

\bibitem[Kriegeskorte et~al., 2009]{Kriegeskorte2009}
Kriegeskorte, N., Simmons, W.~K., Bellgowan, P. S.~F., and Baker, C.~I. (2009).
\newblock Circular analysis in systems neuroscience: the dangers of double
  dipping.
\newblock {\em Nature Neuroscience}, 12(5):535--540.

\bibitem[Kr{\'{o}}l and Kr{\'{o}}l, 2019]{Krol2019}
Kr{\'{o}}l, M. and Kr{\'{o}}l, M.~E. (2019).
\newblock {A novel eye movement data transformation technique that preserves
  temporal information: A demonstration in a face processing task}.
\newblock {\em Sensors}, 19(10):2377.

\bibitem[Kr\"{u}ger et~al., 2017]{Krueger2017}
Kr\"{u}ger, M., Wiebel, C.~B., and Wersing, H. (2017).
\newblock From tools towards cooperative assistants.
\newblock In {\em Proceedings of the 5th International Conference on Human
  Agent Interaction (HAI)}, page 287–294, New York, NY, USA. Association for
  Computing Machinery.

\bibitem[K{\"{u}}bler et~al., 2017]{Kuebler2017}
K{\"{u}}bler, T.~C., Rothe, C., Schiefer, U., Rosenstiel, W., and Kasneci, E.
  (2017).
\newblock {SubsMatch 2.0: Scanpath comparison and classification based on
  subsequence frequencies}.
\newblock {\em Behavior Research Methods}, 49(3):1048--1064.

\bibitem[Lizier et~al., 2012a]{Lizier2012c}
Lizier, J.~T., Prokopenko, M., and Zomaya, A.~Y. (2012a).
\newblock Coherent information structure in complex computation.
\newblock {\em Theory in Biosciences}, 131(3):193--203.

\bibitem[Lizier et~al., 2012b]{Lizier2012}
Lizier, J.~T., Prokopenko, M., and Zomaya, A.~Y. (2012b).
\newblock {Local measures of information storage in complex distributed
  computation}.
\newblock {\em Information Sciences}, 208:39--54.

\bibitem[Lizier and Rubinov, 2012]{Lizier2012b}
Lizier, J.~T. and Rubinov, M. (2012).
\newblock Multivariate construction of effective computational networks from
  observational data.
\newblock {\em Preprint no.: 25/2012, Max Planck Institute for Mathematics in
  the Sciences}.
\newblock Available from:
  \url{https://www.mis.mpg.de/publications/preprints/2012/prepr2012-25.html}
  (accessed: 2020-12-15).

\bibitem[MacKay, 2005]{MacKay2005}
MacKay, D. J.~C. (2005).
\newblock {\em {Information Theory, Inference, and Learning Algorithms}}.
\newblock Cambridge University Press, Cambridge, UK.

\bibitem[Majaranta and Bulling, 2014]{Majaranta2014}
Majaranta, P. and Bulling, A. (2014).
\newblock Eye tracking and eye-based human-computer interaction.
\newblock In {\em Advances in Physiological Computing}, pages 39--65. Springer.

\bibitem[Miller, 1955]{Miller1955}
Miller, G. (1955).
\newblock Note on the bias of information estimates.
\newblock In Quastler, H., editor, {\em Information Theory in Psychology II-B},
  pages 95--100, Glencoe, IL. Free Press.

\bibitem[Novelli et~al., 2019]{Novelli2019}
Novelli, L., Wollstadt, P., Mediano, P. A.~M., Wibral, M., and Lizier, J.~T.
  (2019).
\newblock {Large-scale directed network inference with multivariate transfer
  entropy and hierarchical statistical testing}.
\newblock {\em Network Neuroscience}, 3(3):827--847.

\bibitem[Paninski, 2003]{Paninski2003}
Paninski, L. (2003).
\newblock Estimation of entropy and mutual information.
\newblock {\em Neural Computation}, 15(6):1191--1253.

\bibitem[Panzeri et~al., 2007]{Panzeri2007}
Panzeri, S., Senatore, R., Montemurro, M.~A., and Petersen, R.~S. (2007).
\newblock Correcting for the sampling bias problem in spike train information
  measures.
\newblock {\em Journal of Neurophysiology}, 98(3):1064--1072.

\bibitem[Panzeri and Treves, 1996]{Panzeri1996}
Panzeri, S. and Treves, A. (1996).
\newblock {Analytical estimates of limited sampling biases in different
  information measures}.
\newblock {\em Network: Computation in Neural Systems}, 7(1):87--107.

\bibitem[Peirce, 2007]{Peirce2007}
Peirce, J.~W. (2007).
\newblock Psychopy -- psychophysics software in python.
\newblock {\em Journal of Neuroscience Methods}, 162(1-2):8--13.

\bibitem[Peirce, 2009]{Peirce2009}
Peirce, J.~W. (2009).
\newblock Generating stimuli for neuroscience using psychopy.
\newblock {\em Frontiers in Neuroinformatics}, 10(2):1--8.

\bibitem[{R Core Team}, 2017]{R2017}
{R Core Team} (2017).
\newblock {\em R: A Language and Environment for Statistical Computing}.
\newblock R Foundation for Statistical Computing, Vienna, Austria.

\bibitem[Raptis et~al., 2017]{Raptis2017}
Raptis, G.~E., Katsini, C., Avouris, N., Belk, M., Fidas, C., and Samaras, G.
  (2017).
\newblock {Using eye gaze data {\&} visual activities to infer human cognitive
  styles: Method {\&} feasibility studies}.
\newblock In {\em UMAP 2017 - Proceedings of the 25th Conference on User
  Modeling, Adaptation and Personalization}, pages 164--173, New York, NY.
  Association for Computing Machinery.

\bibitem[Salvucci and Goldberg, 2000]{Salvucci2000}
Salvucci, D.~D. and Goldberg, J.~H. (2000).
\newblock Identifying fixations and saccades in eye-tracking protocols.
\newblock In {\em Proceedings of the 2000 Symposium on Eye Tracking Research
  and Applications (ETRA)}, pages 71--78, New York, NY, USA. ACM.

\bibitem[Sareen et~al., 2016]{Sareen2016}
Sareen, P., Ehinger, K.~A., and Wolfe, J.~M. (2016).
\newblock {CB} database: A change blindness database for objects in natural
  indoor scenes.
\newblock {\em Behavior Research Methods}, 48(4):1343--1348.

\bibitem[Schieber and Gilland, 2008]{Schieber2008}
Schieber, F. and Gilland, J. (2008).
\newblock {Visual entropy metric reveals differences in drivers' eye gaze
  complexity across variations in age and subsidiary task load}.
\newblock {\em Proceedings of the Human Factors and Ergonomics Society},
  3:1883--1887.

\bibitem[Shannon, 1948]{Shannon1948}
Shannon, C.~E. (1948).
\newblock {A mathematical theory of communication}.
\newblock {\em The Bell System Technical Journal}, 27:379--423.

\bibitem[Shiferaw et~al., 2019a]{Shiferaw2019}
Shiferaw, B., Downey, L., and Crewther, D. (2019a).
\newblock A review of gaze entropy as a measure of visual scanning efficiency.
\newblock {\em Neuroscience and Biobehavioral Reviews}, 96:353--366.

\bibitem[Shiferaw et~al., 2019b]{Shiferaw2019alcohol}
Shiferaw, B.~A., Crewther, D.~P., and Downey, L.~A. (2019b).
\newblock Gaze entropy measures detect alcohol-induced driver impairment.
\newblock {\em Drug and Alcohol Dependence}, 204:107519.

\bibitem[Shiferaw et~al., 2018]{Shiferaw2018}
Shiferaw, B.~A., Downey, L.~A., Westlake, J., Stevens, B., Rajaratnam, S.
  M.~W., Berlowitz, D.~J., Swann, P., and E., H.~M. (2018).
\newblock Stationary gaze entropy predicts lane departure events in
  sleep-deprived drivers.
\newblock {\em Scientific Reports}, 8(1):2220.

\bibitem[Simola et~al., 2008]{Simola2008}
Simola, J., Saloj{\"{a}}rvi, J., and Kojo, I. (2008).
\newblock Using hidden {Markov} model to uncover processing states from eye
  movements in information search tasks.
\newblock {\em Cognitive Systems Research}, 9(4):237--251.

\bibitem[Sutton, 1988]{Sutton1988}
Sutton, R.~S. (1988).
\newblock Learning to predict by the methods of temporal differences.
\newblock {\em Machine Learning}, 3(1):9--44.

\bibitem[van Dijk et~al., 2011]{Dijk2011}
van Dijk, H., van~de Merwe, K., and Zon, R. (2011).
\newblock A coherent impression of the pilots' situation awareness: studying
  relevant human factors tools.
\newblock {\em The International Journal of Aviation Psychology},
  21(4):343--356.

\bibitem[Vicente et~al., 2011]{Vicente2011}
Vicente, R., Wibral, M., Lindner, M., and Pipa, G. (2011).
\newblock {Transfer entropy-a model-free measure of effective connectivity for
  the neurosciences}.
\newblock {\em Journal of Computational Neuroscience}, 30(1):45--67.

\bibitem[Wang et~al., 2012]{Wang2012}
Wang, X.~R., Miller, J.~M., Lizier, J.~T., Prokopenko, M., and Rossi, L.~F.
  (2012).
\newblock Quantifying and tracing information cascades in swarms.
\newblock {\em PLOS ONE}, 7(7):e40084.

\bibitem[Wibral et~al., 2014]{Wibral2014}
Wibral, M., Lizier, J.~T., V{\"{o}}gler, S., Priesemann, V., and Galuske, R.
  (2014).
\newblock {Local active information storage as a tool to understand distributed
  neural information processing}.
\newblock {\em Frontiers in Neuroinformatics}, 8:1--11.

\bibitem[Wiebel-Herboth et~al., 2020]{Wiebel-Herboth2020}
Wiebel-Herboth, C.~B., Kr{\"u}ger, M., and Hasenj{\"a}ger, M. (2020).
\newblock Interactions between inter- and intra-individual effects on gaze
  behavior.
\newblock In {\em Adjunct Publication of the 28th ACM Conference on User
  Modeling, Adaptation and Personalization (UMAP)}, pages 35--40, New York, NY.
  Association for Computing Machinery.

\bibitem[Wollstadt et~al., 2019]{Wollstadt2019}
Wollstadt, P., Lizier, J.~T., Vicente, R., Finn, C., Mart\'inez-Zarzuela, M.,
  Mediano, P. A.~M., Novelli, L., and Wibral, M. (2019).
\newblock {IDTxl: The Information Dynamics Toolkit xl: a Python package for the
  efficient analysis of multivariate information dynamics in networks}.
\newblock {\em Journal of Open Source Software}, 4(34):1081.

\bibitem[Wollstadt et~al., 2017]{Wollstadt2017}
Wollstadt, P., Sellers, K.~K., Rudelt, L., Priesemann, V., Hutt, A.,
  Fr{\"o}hlich, F., and Wibral, M. (2017).
\newblock Breakdown of local information processing may underlie isoflurane
  anesthesia effects.
\newblock {\em PLOS Computational Biology}, 13(6):e1005511.

\bibitem[Zagermann et~al., 2018]{Zagermann2018}
Zagermann, J., Pfeil, U., and Reiterer, H. (2018).
\newblock Studying eye movements as a basis for measuring cognitive load.
\newblock In {\em Extended Abstracts of the 2018 Conference on Human Factors in
  Computing Systems (CHI)}, pages 1--6.

\end{thebibliography}

\end{document}